# Evolution of the Fe-3d impurity band state as the origin of high Curie temperature in p-type ferromagnetic semiconductor (Ga,Fe)Sb

Takahito Takeda[1], Shoya Sakamoto[2], Kohsei Araki[1], Yuita Fujisawa[3], Le Duc Anh[1,4], Nguyen Thanh Tu[5], Yukiharu Takeda[6], Shin-ichi Fujimori[6], Atsushi Fujimori[7,8], Masaaki Tanaka[1,9], and Masaki Kobayashi[1,9]

[1]Department of Electrical Engineering and Information Systems, The University of Tokyo, 7-3-1 Hongo, Bunkyo-ku, Tokyo 113-8656, Japan
[2]The Institute for Solid State Physics, The University of Tokyo, 5-1-5 Kashiwanoha, Kashiwa, Chiba 277-8581, Japan
[3]Quantum Materials Science Unit, Okinawa Institute of Science and Technology Graduate University, 1919-1 Tancha, Onna-son, Okinawa, 904-0495, Japan
[4]Institute of Engineering Innovation. The University of Tokyo, 7-3-1 Hongo, Bunkyo-ku, Tokyo 113-0032, Japan
[5]Department of Physics, Ho Chi Minh City University of Pedagogy, 280, Au Duong Vuong Street, District 5, Ho Chi Minh City 748242, Vietnam
[6]Materials Sciences Research Center, Japan Atomic Energy Agency, Sayo-gun, Hyogo 679-5148, Japan
[7]Department of Physics, The University of Tokyo, 7-3-1 Hongo, Bunkyo-ku, Tokyo 113-0033, Japan
[8]Department of Applied Physics, Waseda University, Okubo, Shinjuku, Tokyo 169-8555, Japan
[9]Center for Spintronics Research Network, The University of Tokyo, 7-3-1 Hongo, Bunkyo-ku, Tokyo 113-8656, Japan

**ABSTRACT**

$(Ga_{1-x},Fe_x)Sb$ is one of the promising ferromagnetic semiconductors for spintronic device applications because its Curie temperature ($T_C$) is above 300 K when the Fe concentration $x$ is equal to or higher than ~0.20. However, the origin of the high $T_C$ in (Ga,Fe)Sb remains to be elucidated. To address this issue, we use resonant photoemission spectroscopy (RPES) and first-principles calculations to investigate the $x$ dependence of the Fe 3d states in $(Ga_{1-x},Fe_x)Sb$ ($x$ = 0.05, 0.15, and 0.25) thin films. The observed Fe 2p-3d RPES spectra reveal that the Fe-3d impurity band (IB) crossing the Fermi level becomes broader with increasing $x$, which is qualitatively consistent with the picture of double-exchange interaction. Comparison between the obtained Fe-3$d$ partial density of states and the first-principles calculations suggests that the Fe-3$d$ IB originates from the minority-spin ($\downarrow$) $e$ states. The results indicate that enhancement of the interaction between $e_\downarrow$ electrons with increasing $x$ is the origin of the high $T_C$ in (Ga,Fe)Sb.

# I. INTRODUCTION

Ferromagnetic semiconductors (FMSs) are alloy semiconductors which show both semiconducting and ferromagnetic properties. In III-V FMSs, cation sites (group III sites) are partially replaced by a sizable amount of magnetic impurities such as Mn and Fe ions. The ferromagnetism of FMSs is considered to be caused by the magnetic interaction between the doped magnetic ions mediated by the spin of the carriers. This ferromagnetism is called carrier-induced ferromagnetism[1]. In order to explain the ferromagnetism of FMSs, two models in which ferromagnetic interaction is mediated by itinerant carriers (band conduction)[2,3] and localized carriers (hopping conduction)[4,5] have been proposed so far. The former and the latter are called the Zener's $p$-$d$ exchange model and the impurity band (IB) model[6], respectively. Here, the characteristic difference between these models is caused by the position of the 3$d$ impurity band relative to the Fermi level in the valence-band structure.

The Mn-doped III-V FMSs, such as (In,Mn)As[7,8,9,10] and (Ga,Mn)As[11,12,13], have been intensively studied as prototypical FMSs for more than twenty years. Nevertheless, they have some issues to be solved for device applications, that are, the Mn-doped FMSs show low Curie temperature ($T_C$) below room temperature[14,15] and have only p-type carriers. Recently, Fe-doped III-V FMSs such as n-type (In,Fe)As[16,17,18], n-type (In,Fe)Sb[19,20,21], and p-type (Ga,Fe)Sb[22,23,24] have been successfully grown for the first time by molecular beam epitaxy (MBE). Since the doped Fe ions in the III-V semiconductors are expected to substitute for the cation ($In^{3+}$ or $Ga^{3+}$) sites as $Fe^{3+}$ and conducting carriers can be introduced by donor/acceptor doping, off-stoichiometry, and defects, one can independently control the concentrations of Fe ions and carriers in Fe-doped FMSs. Furthermore, the highest $T_C$ values reported so far in $(In_{0.65},Fe_{0.35})Sb$ (385 K)[21] and $(Ga_{0.8},Fe_{0.2})Sb$ (> 400 K)[25] are well above room temperature. Considering these advantages, Fe-based FMSs are more promising materials for practical semiconductor spintronics devices operating at room temperature. However, the origin of the high $T_C$ has not been unveiled yet. Since the electronic structures of Fe-doped FMSs with $T_C$ higher than room temperature have not been fully clarified, understanding the origin of the high-$T_C$ ferromagnetism from the electronic structure point of view is indispensable for the design and realization of functional FMSs.

As for (Ga,Fe)Sb, some experimental[26,27,28] and theoretical[29,30,31] studies on its electronic structure were performed to understand the mechanism of the carrier-induced ferromagnetism. The local electronic structure, magnetic properties of Fe ions, and the valence-band (VB) structure were investigated by x-ray absorption spectroscopy (XAS), x-ray magnetic circular dichroism (XMCD), and resonant photoemission spectroscopy (RPES)[26]. From these experiments, it was found that Fe ions in (Ga,Fe)Sb take the $3d^6$ configuration, which has both itinerant and correlated nature of $3d$ electrons, and there is a finite Fe partial density of states (PDOS) at the Fermi level ($E_F$). The Fe PDOS at $E_F$ is consistent with the IB model and the itinerant nature of $3d$ electrons is attributed to the $p$-$d$ exchange interaction through the finite s$p$-$d$ hybridization. Hence these results suggest that both the double-exchange interaction and the $p$-$d$ exchange interaction contribute to the ferromagnetism. We have recently studied the electronic states of ($Ga_{0.95}$,$Fe_{0.05}$)Sb in the vicinity of $E_F$ by soft x-ray angle-resolved photoemission spectroscopy (SX-ARPES)[27]. This study indicates that the electronic structure of ($Ga_{0.95}$,$Fe_{0.05}$)Sb is consistent with the IB model and the origin of its ferromagnetism is the double-exchange interaction. The first-principles calculations for (Ga,Fe)Sb[29,30,31] well reproduce the observed electronic structure. The Fe-concentration dependence of the VB of (Ga,Fe)Sb has been studied by infrared (IR) magnetic circular dichroism (MCD)[28]. The IR-MCD results can be explained by the IB model and suggest that $E_F$ gradually move upwards with the increase of the Fe concentration. However, the origin of the high $T_C$ in (Ga,Fe)Sb is still poorly understood, since the electronic structure of the heavily-Fe-doped (Ga,Fe)Sb has never been directly observed so far. In this article, we reveal the Fe-concentration ($x$) dependence of the Fe 3$d$ states of ($Ga_{1-x}$,$Fe_x$)Sb with $x$ = 0.05, 0.15, and 0.25 by RPES measurements. Systematic analyses of the Fe 2$p$-3$d$ RPES spectra indicate that the Fe 3$d$ states near $E_F$ play a key role to induce the high-$T_C$ ferromagnetism.

## II. EXPERIMENTAL

($Ga_{1-x}$,$Fe_x$)Sb thin films ($x$ = 0.05, 0.15, and 0.25) with a thickness of 30 nm were grown on p-type GaAs(001) substrates by MBE, where the growth condition was the same as the previous research[22,23]. The surface of the films was covered by a thin amorphous Sb capping layer to avoid surface contamination. The sample structure is, from top to bottom, Sb capping layer 1-2 nm/($Ga_{1-x}$,$Fe_x$)Sb 30 nm/GaSb 100 nm/AlSb 100 nm/AlAs 10nm/GaAs 100 nm/p-GaAs substrate. Since the thickness of the capping layer was comparable with the mean free path or escape depth (1 ~ 2 nm) of photoelectrons in the soft x-ray region, we were able to measure photoemission signals without removing the capping layer[32]. During the MBE growth, the excellent crystallinity of the samples was confirmed by reflection high-energy electron diffraction. The values of $T_C$ of ($Ga_{1-x}$,$Fe_x$)Sb with $x$ = 0.05, 0.15, and 0.25, which were estimated by MCD[22,23,24], are about 40 K, 125 K, and above 300 K, respectively. The RPES experiments were performed at beamline BL23SU of SPring-8. The measurements were conducted under an ultrahigh vacuum below $1.3 \times 10^{-8}$ Pa at a temperature of 30 K. The photon energy ($hv$) of the incident beam was varied from 690 eV to 740 eV and circular polarization was used for the measurements. The total energy resolution including the thermal broadening was between 100 meV and 160 meV depending on $h\nu$. The $E_F$ position of RPES spectra has been corrected by the Fermi edge of an Au foil in electrical contact with the samples. Fe $L_{2,3}$ x-ray absorption spectroscopy (XAS) spectra were measured in the total-electron-yield mode.

First-principles calculations were performed on a 3×3×3 GaSb supercell with one substitutional Fe atom, namely, $(Ga_{26},Fe_1)Sb_{27}$ (3.7% Fe), using the full-potential augmented-plane-wave method as implemented in the WIEN2k code[33]. The experimental lattice constant of 6.085 Å, estimated from Vegard's law in Ref. 23, was used for the calculation, and lattice relaxation was not taken into account. For the exchange-correlation potential, the generalized gradient approximation (GGA) of Perdew-Burke-Erzerhof parametrization[34] was employed with/without the empirical Hubbard $U$ (GGA+$U$)[35] accounting for the on-site Coulomb correlation of Fe $3d$ electrons. Spin-orbit interaction was also included. The Brillouin-zone integration was performed on a 4 x 4 x 4 $k$-point mesh.

### III. RESULTS AND DISCUSSION

## A. XAS spectra at the Fe $L_{2,3}$ absorption edge

Figure 1(a) shows XAS spectra at the Fe $L_{2,3}$ absorption edge of $(Ga_{1-x},Fe_x)Sb$ ($x$ = 0.05, 0.15, and 0.25) (black curves), where black dashed curves represent the background components. Here, the Fe $L_{2,3}$ XAS spectrum of α-$Fe_2O_3$ is also shown as a reference spectrum of $Fe^{3+}$ (blue curve)[36]. Besides the main peak at ∼ 708 eV, there is a shoulder structure at ∼ 710 eV in the XAS spectra of (Ga,Fe)Sb. The previous XMCD and RPES studies[26] showed that the shoulder component can be attributed to $Fe^{3+}$ oxides (such as α-$Fe_2O_3$) and is not related to the ferromagnetism in $(Ga_{1-x},Fe_x)Sb$. The oxidized components in the present capped (Ga,Fe)Sb samples are less than that in a previous uncapped (Ga,Fe)Sb thin film[26], suggesting that the surface of our samples is well protected by the amorphous Sb capping layer from oxidation.

To extract the intrinsic components of (Ga,Fe)Sb, the Fe $L_{2,3}$ XAS spectra have been decomposed using the $Fe^{3+}$ reference spectrum. Figures 1(b)-1(d) show the decomposition analysis for the XAS spectra of $(Ga_{1-x},Fe_x)Sb$. Here, the red, black, and blue curves represent intrinsic XAS spectra of (Ga,Fe)Sb, raw XAS spectra, and the reference spectra of $Fe^{3+}$, respectively. The intrinsic XAS spectra are obtained by eliminating the Fe-oxide component from the raw XAS spectra. Comparing these intrinsic spectra, the line-shapes of the spectra are nearly identical to each other irrespective of $x$. This result indicates that the valence states (local electronic structure) of Fe ions are independent of $x$ in (Ga,Fe)Sb.

**B. RPES: Comparison between on- and off-resonance spectra**

According to the previous SX-ARPES result[27], the electronic state of Fe $3d$ in the vicinity of $E_F$ plays a key role in the ferromagnetism based on the double-exchange interaction, which originates from the exchange of electrons between Fe ions through Sb ions. First-principles calculations suggest that the IB broadens with increasing the concentration of magnetic impurities in the case of double-exchange interaction due to kinetic energy gain[37,38]. Thus, the VB structure of (Ga,Fe)Sb is expected to depend on $x$. To characterize the $x$ dependence of the Fe-$3d$ PDOS in the VB structure, we have conducted RPES measurements on the $(Ga_{1-x},Fe_x)Sb$ ($x = 0.05$, $0.15$, and $0.25$) thin films and analyzed the spectra in detail. The resonant enhancement of the photoemission intensity measured at $hv = 708$ eV should reflect the energy position of the ferromagnetic component of Fe 3$d$ states in (Ga$_{1-x}$,Fe$_x$)Sb, because Fe $L_3$ XMCD has a prominent peak at $hv$ = 708 eV, the intensity of which as a function of magnetic field shows a ferromagnetic behavior[26]. Figure 2(a) shows the RPES spectra of the (Ga$_{1-x}$,Fe$_x$)Sb thin films taken at $hv$ = 708 eV (on-resonance) and $hv$ = 704 eV (off-resonance), respectively. These spectra are normalized to the intensity of incident x rays. To extract the resonant behavior, the off-resonance spectra are subtracted from the on-resonance ones. Figure 2(b) shows the differences between the on- and off-resonance spectra. Depending on $x$, spectral differences appear at around $E_F$ (black dashed arrows) and binding energy ($E_B$) of 1.6 eV (black solid arrows). That is, the intensity at around $E_F$ increases and the peak at around 1.6 eV becomes vague with increasing $x$.

The previous RPES study of (Ga,Fe)Sb[26] demonstrated that the difference spectra are composed of symmetric and asymmetric Gaussian components originating from the Fe-3$d$ PDOS and Auger peak, respectively. In the previous RPES study, there were two symmetric components around $E_B$ = 1.7 eV and 10.3 eV and an asymmetric Auger component around $E_B$ = 4.0 eV. To identify these components in the present study, the difference spectra are fitted by a linear combination of the exponentially modified Gaussian for the asymmetric Auger component, which do not represent the Fe-3$d$ electronic states, and Gaussian functions for the symmetric component, as shown in Figs. 3(a)-3(c). These fittings are conducted except for the range in the vicinity of $E_F$ (-0.14 eV – 1.2 eV). The symmetric components around $E_B = 1.6$ eV and $E_B = 2.6$ eV are denoted by α and γ, respectively. No peak around $E_B = 2.6$ eV has been observed in the chemically etched (Ga,Fe)Sb[26]. Since the γ component is not resonantly enhanced at $hv = 710$ eV, which corresponds to the main peak of the $Fe^{3+}$ XAS spectrum, the γ component does not originate from the oxidized $Fe^{3+}$ ions. Red curves shown in Fig. 3(d)- 3(f) represent the spectra after subtracting the fitted asymmetric Gaussian component from the difference spectra, which reflect the PDOS of Fe 3$d$ electrons. The difference between the Fe-PDOS spectra and the α + γ spectra corresponds to the Fe-3$d$ IB in the vicinity of $E_F$ observed with SX-ARPES[27], as shown by blue filled areas in Fig. 3 (d) - (f).

**C. Fe concentration dependence of Fe-3d PDOS**

The peak position and the width of the α component depend on $x$. To elucidate the $x$ dependence of the α and γ components quantitatively, the values of peak positions (PP) and full width at half maximum (FWHM) of α and γ for $(Ga_{1-x},Fe_x)Sb$ are compared. The values of FWHM represent the broadening of the states. Figure 4(a) shows the PP of the α (red markers) and γ (blue markers) components relative to that of $(Ga_{0.95},Fe_{0.05})Sb$ as a function of $x$. The value of PP of the α component increases with increasing $x$ as shown in Fig.4(a). In contrast, the PP of the γ component seems independent of $x$. Figure 4(b) shows the FWHM of the α and γ components. While the value of FWHM of the γ component is nearly identical irrespective of $x$ within the experimental accuracy, that of the α component increases with increasing $x$. Both the PP and FWHM of the γ components are independent of $x$, indicating that the γ is an extrinsic component. As described above, the γ component is different from the $Fe^{3+}$ oxides, and therefore, the γ component likely originates from some kinds of extrinsic precipitates between the surface of the $(Ga_{1-x},Fe_x)Sb$ films and the capping layers. On the other hand, the PP and FWHM of the α component systematically change with $x$, reflecting the change of the intrinsic Fe-3$d$ PDOS with $x$ in $(Ga_{1-x},Fe_x)Sb$.

The IR-MCD study on $(Ga_{1-x},Fe_x)Sb$[28] has demonstrated that the *sp* bands move to the high-$E_B$ direction with the increase of $x$ because the $E_F$ moves upwards. The shifts of the *sp* bands relative to those of $(Ga_{0.95},Fe_{0.05})Sb$ estimated from the IR-MCD results are plotted in Fig. 4(a) (purple markers). Figure 4(a) shows that the $x$ dependence of the PP of the α component follows the same trend as that of the *sp* bands. This result indicates that the shift of the α component with increasing $x$ corresponds to the chemical potential ($E_F$) shift in (Ga,Fe)Sb.

The $x$ dependence of the Fe-3$d$ IB near $E_F$ is different from that of the α component. Figure 5(a) shows the intrinsic Fe-3$d$ PDOS with different $x$. It should be noted here that the shifts of the Fe-3$d$-IB tail (solid blue arrows) with increasing $x$ are larger than the peak shifts of the α component (black dashed arrows). This indicates that the Fe-3$d$ IB broadens with increasing $x$ since the difference of the shifts is not explained by the chemical potential shift. The broadening of the Fe-3$d$ IB is also suggested in the IR-MCD study of (Ga,Fe)Sb[28]. The broadening of the IB with the increase of magnetic impurities is also predicted by first-principles calculations for (Ga,Mn)N[38], where the Mn-3$d$ IB crosses $E_F$ as in the case of Fe-3$d$ IB in (Ga,Fe)Sb. This broadening of the Fe-3$d$ IB with increasing $x$ observed in (Ga,Fe)Sb is qualitatively consistent with the picture of double-exchange interaction[38,39,40]. Therefore, this result indicates that the origin of ferromagnetism in (Ga,Fe)Sb is double-exchange interaction irrespective of $x$.

**D. Discussion**

Based on the experimental findings, let us discuss the $x$ dependence of the Fe-3$d$ electronic states and its effect on the ferromagnetism. From the observations described above, with the increase of $x$, the Fe 3$d$-IB component below $E_\mathrm{F}$ and the α peak become broader and the PP of α moves to the high-$E_\mathrm{B}$ direction.

To discuss the origin of the α component, the $p$-$d$ hybridization should be taken into account. The strength of the $p$-$d$ hybridization is significantly affected by the symmetry of $d$ electrons in (Ga,Fe)Sb. The five-fold degenerate states of Fe 3$d$ orbitals split into the two-fold degenerate e states and the three-fold degenerate $t_2$ states, because the Fe ions of (Ga,Fe)Sb are in the tetrahedral crystal field. Because of the symmetries of the $e$ and $t_2$ states, while the $e$ states do not hybridize strongly with the ligand $p$ bands, the $t_2$ states do hybridize strongly with these bands[41]. The $p$-$d(t_2)$ hybridization leads to the antibonding ($t_{2a}$) and bonding ($t_{2b}$) states, which have both the Fe $t_2$ and the ligand Sb $p$ characters.

Figure 6(a) shows the results of the first-principles calculations for spin-resolved density of states (DOS) of (Ga,Fe)Sb, where the total DOS, the PDOS of the Fe $e$ states, and the PDOS of the $t_2$ states are plotted by gray area, green curve, and blue curve, respectively. The Fe 3$d$ PDOS of majority-spin (↑) at around $E_B$ = 1.5 eV is much smaller than that of minority-spin (↓). Comparing with the observation and the calculation, the Fe PDOS measured at around $E_B$ = 1.5 eV likely comes from the broad minority-spin states, suggesting that the α component corresponds to the $t_{2b\downarrow}$ state. As discussed above, the $x$ dependence of the PP of $t_{2b\downarrow}$ is the same as that of the $sp$ bands, as shown in Fig. 4(a). Since the Fe-3$d$ IB is located near $E_F$, the Fe-3$d$ IB are gradually occupied with the increase of $x$. Here, this Fe-3$d$ IB near $E_F$ is mainly the minority-spin state[27]. The gradual occupation of the Fe-3$d$ IB by electrons is related to the broadening of the hybridized states with increasing $x$ as shown in Fig. 4(b), the $x$ dependence of the FWHM of the α indicates that the $p$-$d(t_2)$ hybridized states broaden with increasing $x$.

The calculation shown in Fig. 6(a) suggests that both the majority-spin $t_{2a\uparrow}$ and the minority-spin $e_\downarrow$ states cross $E_F$. Considering the narrow band width shown in Fig. 5(a), the Fe-3$d$ IB likely originates from the minority-spin $e_\downarrow$ state.

In the previous SX-ARPES measurements on $(Ga_{0.95},Fe_{0.05})Sb$[27], the origin of the Fe-3$d$ IB in the vicinity of $E_F$ remained experimentally unclear, although it was expected that the Fe-3$d$ IB consists of the $e_\downarrow$ or $t_{2a\downarrow}$ states. To identify the origin of the IB state, we have elucidated the Fe-concentration dependence of the Fe-3$d$ IB near $E_F$. Figure 5(b) shows the enlarged plot of the intrinsic Fe-3$d$ PDOS near $E_F$. Note that the slopes in the vicinity of $E_F$ of the intrinsic Fe-PDOS spectra (Fe-3$d$ IB) are slightly different from the Fermi-Dirac distribution curves (green curve). In particular, the distance between the red and green triangles at $E_F$ becomes larger as $x$ increases. The observation suggests either pseudogap or depletion of DOS near $E_F$ possibly occurs in the Fe PDOS at $E_F$, where pseudogap means the decrease of DOS near $E_F$ caused by gap opening. Additionally, the increase of the distance between the red and green triangles at $E_F$ with the increase of $x$ in Fig. 5(b) suggests that the decrease of the PDOS near $E_F$ becomes large. It is likely that this pseudogap or DOS depletion reflects the localized character of the Fe-3$d$ state composing the Fe-3$d$ IB. Since the $e_\downarrow$ states are more localized than the hybridized $t_{2a,b}$ states, the Fe-3$d$ IB predominantly consists of the $e_\downarrow$ states. The $d$-$d$ Coulomb interaction and/or the degradation of the crystallinity with the increase of $x$ are supposed to induce the pseudogap or depletion of DOS near $E_\text{F}$ as discussed in the following. The pseudogap behavior is hardly seen in the calculation shown in Fig. 6(a). To qualitatively elucidate the effect of the Coulomb interaction $U$ on the Fe-3$d$ IB, first-principle calculations taking into account $U$ (= 2 eV) have been performed as shown in Fig. 6(b). The calculation with $U$ suggests that the energy gap of the $e_\downarrow$ states will open, while the delocalized $t_{2a\uparrow}$ states cross $E_\text{F}$ even under the finite $U$. Actual $U$ in (Ga,Fe)Sb is considerably smaller than the value (2 eV) assumed in the calculation shown in Fig. 6(b), because the gap is not fully opened in the observed spectra and the Fe 3$d$ PDOS is approximately reproduced by the calculation without $U$ shown in Fig. 6(a). When the $e_\downarrow$ state is close to half filled with the increase of $x$, the effect of the Coulomb interaction will increase. Additionally, as for the deterioration of the crystal quality, a theoretical study of computational many-body methods for manganite La$_{1-x}$Sr$_x$MnO$_3$ suggests that the pseudogap features appear at $E_\text{F}$ when the system contains ferromagnetic cluster regions in an insulating background[42]. Since the Fe distribution in heavily-Fe-doped (Ga,Fe)Sb is reportedly nonuniform[43], the inhomogeneity will lead to the pseudogap formation at $E_F$. Then, the observed pseudogap behavior likely reflects the effect of $U$ on the $e_\downarrow$ state and/or the nonuniform distribution of the Fe ions. It follows from these arguments that the Fe-3$d$ IB in the vicinity of $E_F$ originates from the $e_\downarrow$ states.

    Based on the findings, the $x$ dependence of the Fe PDOS per Fe atom is schematically shown in Fig. 7. When $x$ increases, the overlap between the majority-spin $t_{2a\uparrow}$ states and the Fe-3$d$ IB states is expected to become larger due to the band broadening. This means that the Fe-3$d$ IB states will be gradually occupied by electrons when $x$ increases, as shown in Fig. 7. Since the minority-spin $e_\downarrow$ is localized, the majority-spin $t_{2a\uparrow}$ predominantly contributes to the electrical conductivity. In contrast to the $t_2$ hybridized IB states related to the ferromagnetism in (Ga,Mn)As, from the electronic structures shown in Fig. 7, the $e_\downarrow$ electrons originating the Fe-3$d$ IB would predominantly contribute to the ferromagnetism in (Ga,Fe)Sb. The double-exchange interaction is considered to be stronger when the number of $d$ electrons mediating the

interaction increases. The broadening of the $e_↓$ state accompanied by the electron occupation of the state with increasing $x$ is consistent with the picture of double-exchange interaction. Since the $d$ electrons are magnetically coupled through double-exchange interaction, the value of $T_C$ is positively correlated with $x$ in (Ga,Fe)Sb at least for $x < 0.25$. On the other hand, in the case of p-type FMS (Ga,Mn)As, the Mn-3$d$ IB in the vicinity of $E_F$ originates from the $p$-$d$ hybridized states[44]. In (Ga$_{1-x}$,Mn$_x$)As, the $p$-$d$ exchange interaction increases with increasing hole-carrier concentration that is proportional to $x$, whereas the increase of $x$ (> 15%) will lead to disordering of the crystallinity and the $sp$ band. As a consequence, the value of $T_C$ will peak out when $x$ increases in (Ga,Mn)As. Therefore, from these arguments, the enhancement of double-exchange interaction in (Ga,Fe)Sb possibly results from the fact that the Fe-3$d$ IB at $E_F$ originates from the $e_↓$ states. The electron occupation of the $e_↓$ state increases with the increase of $x$, as shown in Fig. 7; this leads to the high $T_C$ observed in heavily doped (Ga$_{1-x}$,Fe$_x$)Sb thin films.

## IV. SUMMARY

We have performed RPES measurements on $(Ga_{1-x},Fe_x)Sb$ ($x$ = 0.05, 0.15, and 0.25) thin films to unveil the origin of the high-$T_C$ ferromagnetism in (Ga,Fe)Sb. The differences between the on- and off-resonance spectra can be decomposed into the intrinsic Fe $3d$ PDOS and extrinsic Fe components. The intrinsic Fe $3d$ PDOS are composed of the *p-d* hybridized state (the α component) in VB and the Fe-$3d$ IB located in the vicinity of $E_F$. The PP of α shifts to the high $E_B$ direction with increasing $x$, indicating that the chemical potential shift depending on $x$ occurs in (Ga,Fe)Sb. The Fe-$3d$ IB near $E_F$ is gradually broadened with increasing $x$. This broadening is consistent with the description of double-exchange interaction. The first-principles calculations reveal that the α and Fe-$3d$ IB originate from the $t_{2b\downarrow}$ and the $e_\downarrow$ states, respectively. Based on the findings, it is concluded that the double-exchange interaction derived from the $e_\downarrow$ electrons is the mechanism of the ferromagnetism in $(Ga_{1-x},Fe_x)Sb$ with $x \leq 0.25$, and the increase of the interaction with increasing the $e_\downarrow$ electrons at $E_F$ would be the origin of the high-$T_C$ ferromagnetism in (Ga,Fe)Sb.

**ACKNOWLEWDGMENTS**

The authors would like to thank K. Sato and H. Katayama-Yoshida for fruitful discussions. This work was supported by Grants-in-Aid for Scientific Research (No.

15H02109, No. 16H02095, No. 19K21961, No. 18H05345, and No. 23000010). L. D. A. acknowledge the support from PRESTO Program (JPMJPR19LB) of Japan Science and Technology Agency. This work was partially supported by the Spintronics Research Network of Japan (Spin-RNJ). This work was performed under the Shared Use Program of Japan Atomic Energy Agency (JAEA) Facilities (Proposal No. 2017B-E19 and 2019A-E15) supported by JAEA Advanced Characterization Nanotechnology Platform as a program of "Nanotechnology Platform" of the Ministry of Education, Culture, Sports, Science and Technology (MEXT) (Proposal No. A-17-AE-0038 and A-19-AE-0015). Supporting experiments at SPring-8 were approved by the Japan Synchrotron Radiation Research Institute (JASRI) Proposal Review Committee (Proposal No. 2017B3841 and No. 2019A2841).

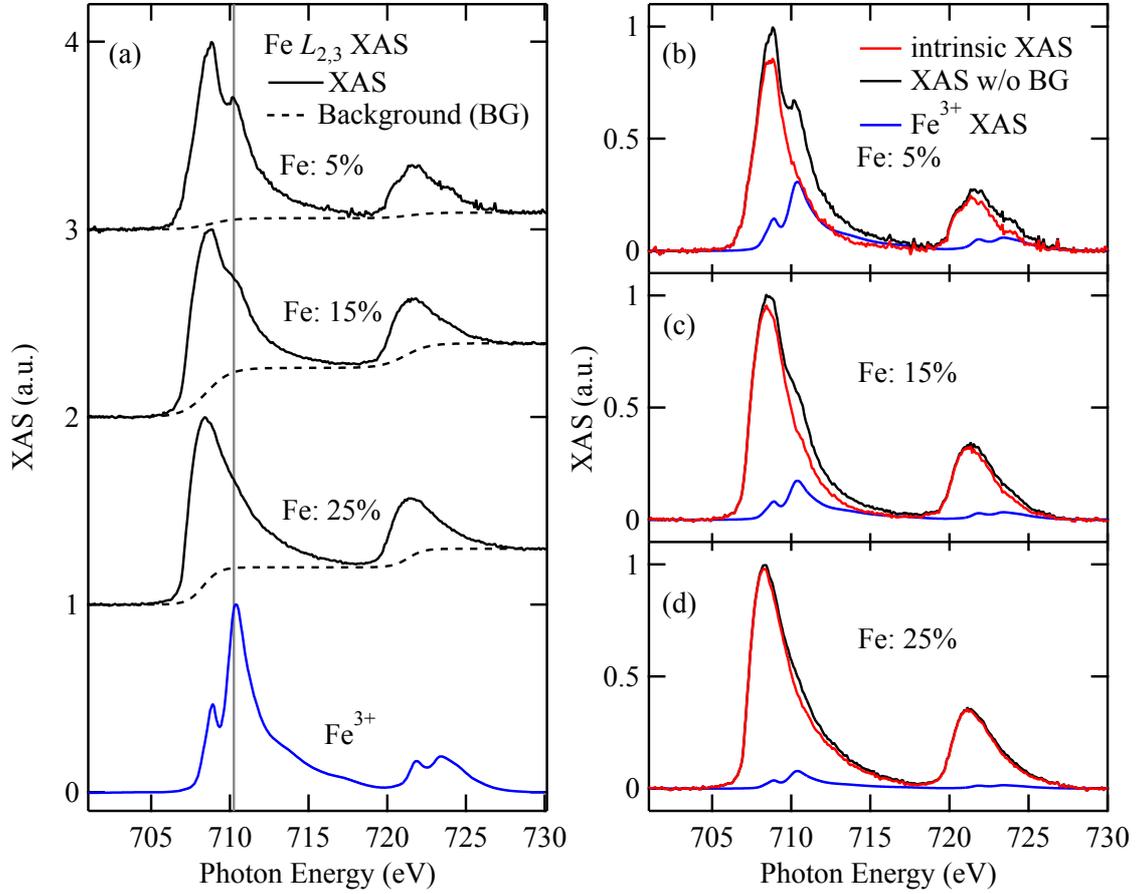

FIG. 1. XAS spectra at the Fe $L_{2,3}$ absorption edge. (a) XAS spectra of $(Ga_{1-x},Fe_x)Sb$ ($x$ = 0.05, 0.15, and 0.25) (black curves) and $Fe^{3+}$ (blue curve) [36]. Black dashed curves represent the backgrounds. Gray vertical line denotes peak position of the $Fe^{3+}$. (b)-(d) Decomposition analysis for the XAS spectra. Intrinsic and extrinsic $Fe^{3+}$ components are separately shown by red and blue curves, respectively. Here, the black curves are the XAS spectra without background in Fig. 1(a).

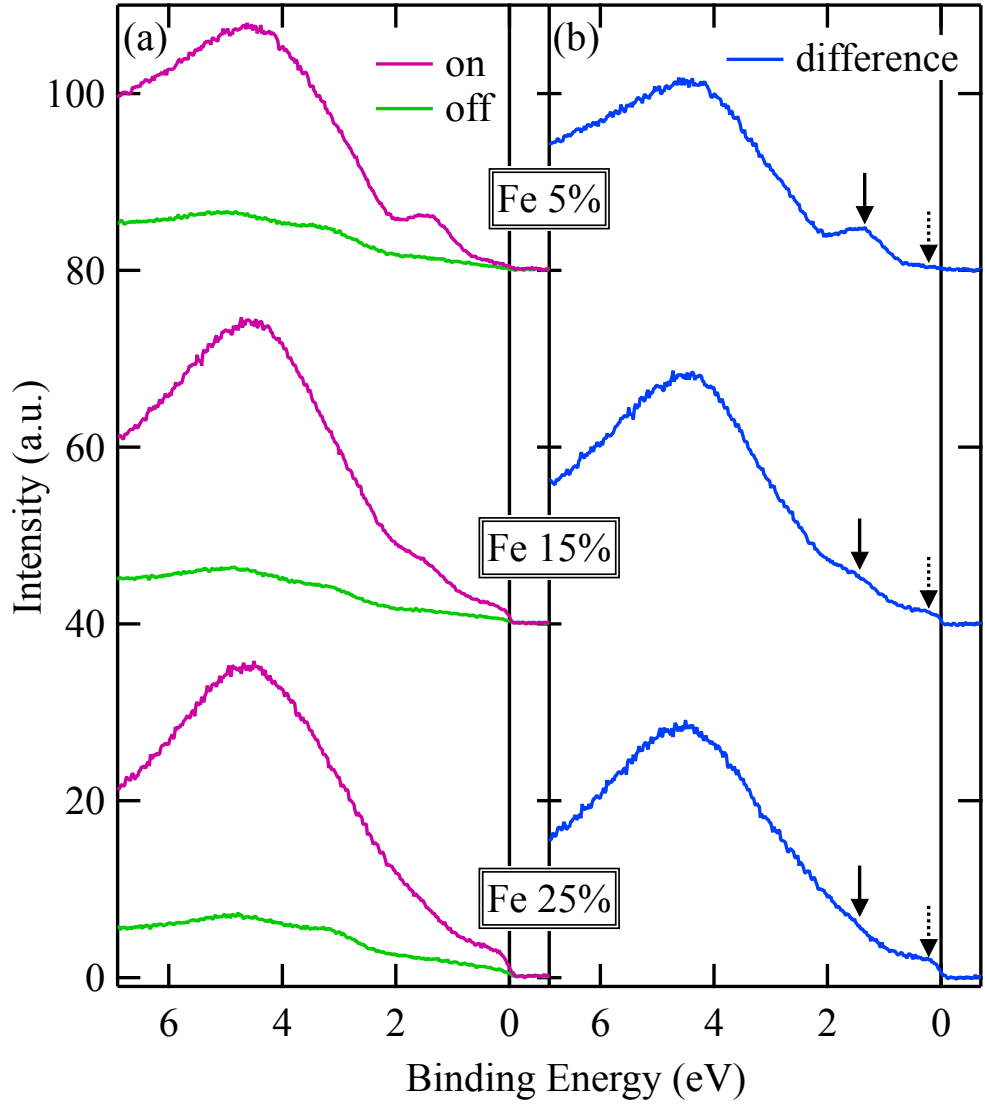

FIG. 2: RPES spectra of $(Ga_{1-x},Fe_x)Sb$ (a) On(off)-resonant spectra of $x$ = 0.05, 0.15, and 0.25 taken at $h\nu$ = 708 eV (704 eV). Purple (green) curves represent on(off)-resonant spectra. (b) The difference spectra (blue curves), which are differences between the on- and off-resonant spectra.

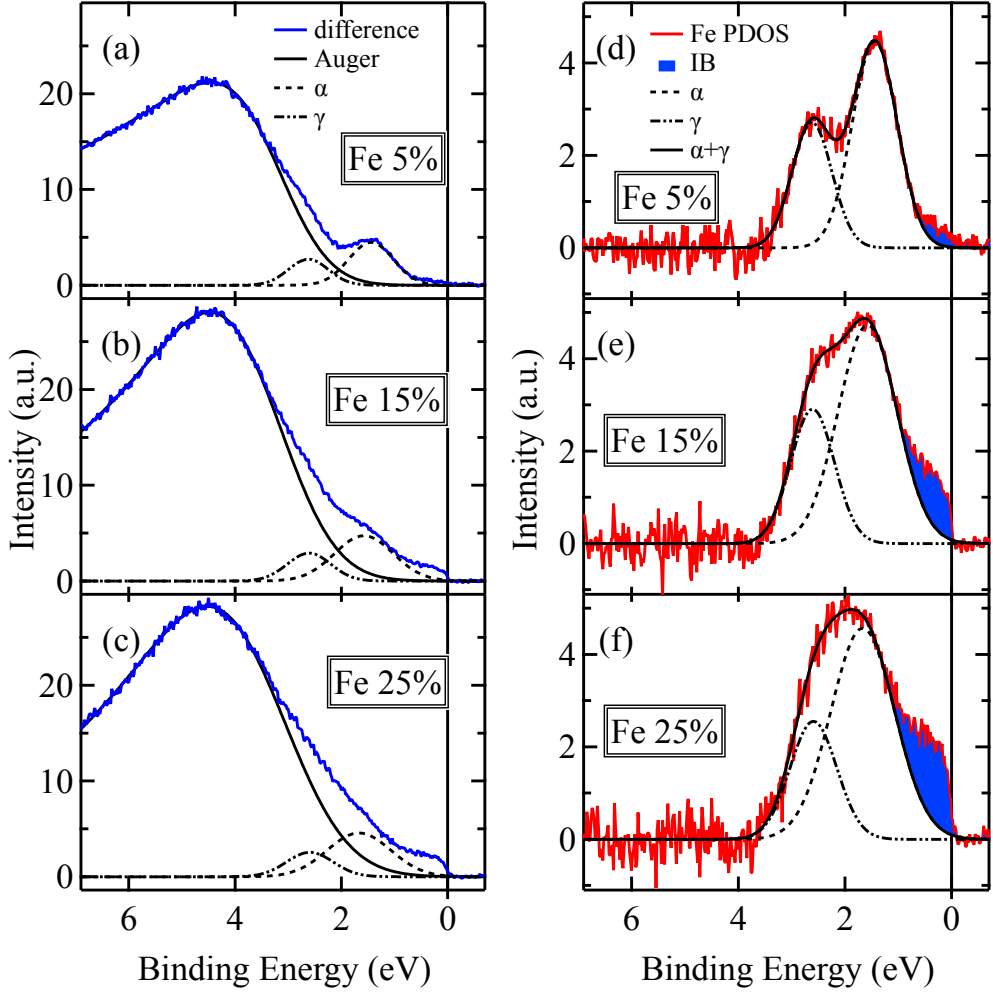

FIG. 3. Decomposition analysis of the RPES spectra of the $(Ga_{1-x},Fe_x)Sb$ thin films. (a)-(c) Subtraction of the Auger component from the spectra shown in Fig. 2(d)-2(f), respectively. Solid black, dashed, dash-dotted, and blue solid curves are Auger components fitted by using asymmetric Gaussian function, fitting components of two Gaussian functions for α and γ components, and the difference spectra. (d)-(f) Fe-PDOS spectra (red curves) are obtained by subtracting the Auger components from the difference spectra (blue curves). Here, dashed, points-dashed, and solid black curves represent fitted α, γ, and α + γ components, respectively. The blue filled areas represent the Fe-3$d$ IB component.

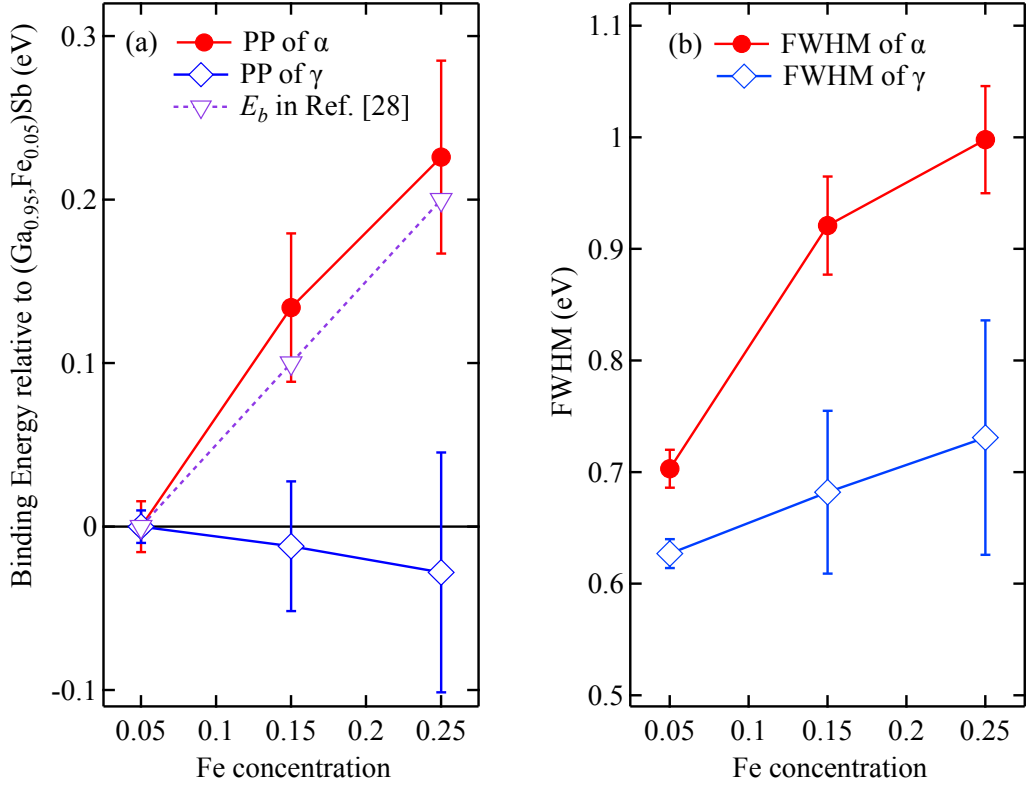

FIG. 4. Peak position (PP) and FWHM of the α and γ components. (a) Binding energy of PP relative to that of $(Ga_{0.95},Fe_{0.05})Sb$ of the α (red markers) and γ (blue markers). Here, $E_b$ is the distance between $E_F$ and the VB at the L point (purple markers). (c) FWHM of the α and γ components. Red and blue points/lines denote the α and γ components, respectively.

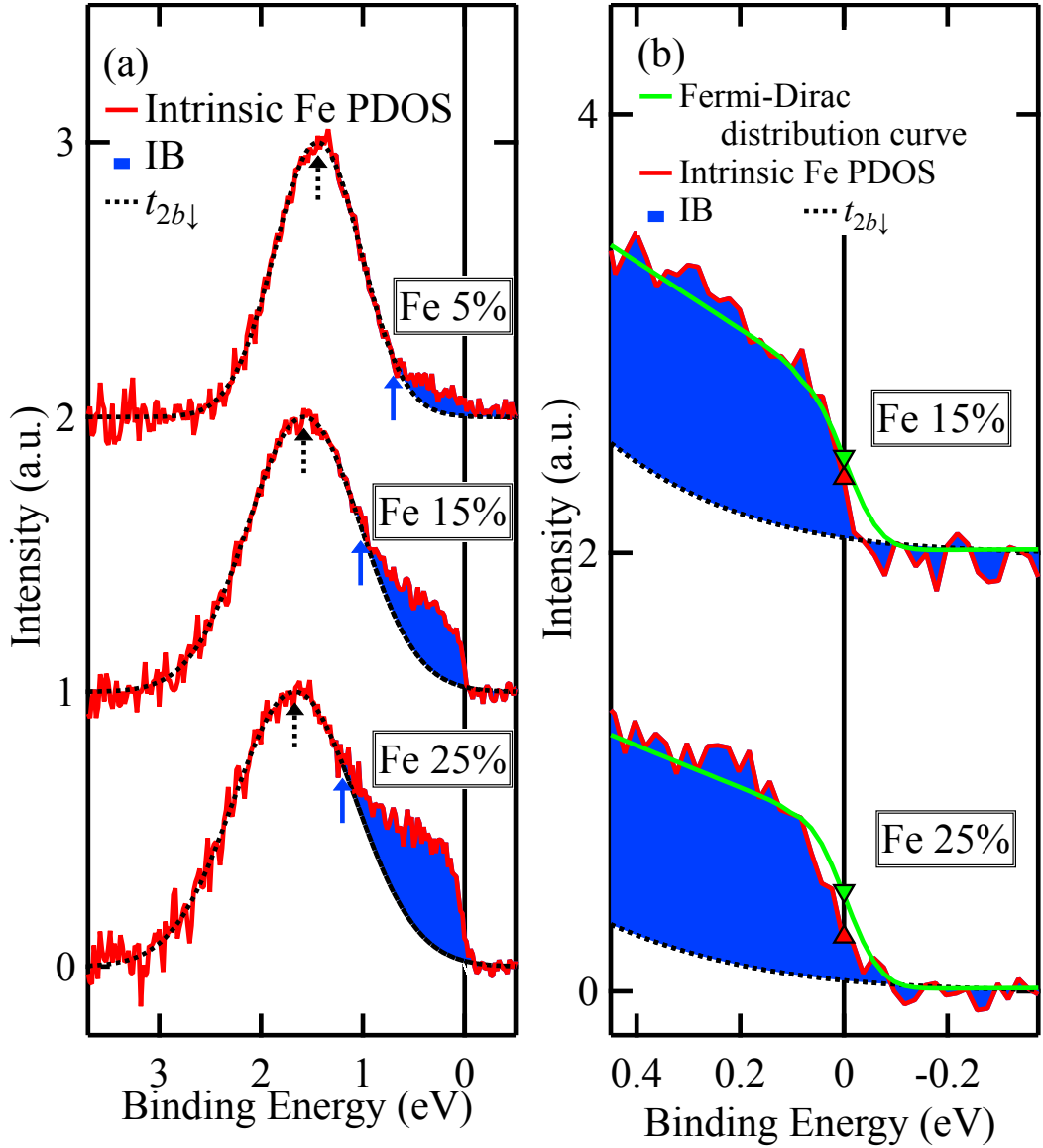

FIG. 5: Intrinsic Fe 3d PDOS of $(Ga_{1-x},Fe_x)Sb$. (a) Spectra with the extrinsic component subtracted from the spectra shown in Fig. 3(d)-3(f). Here, dashed and solid red curves represent fitted α, and intrinsic Fe 3d-PDOS components, respectively. The dashed black arrows denote the peak position of $t_{2b\downarrow}$. The blue solid arrows are a guide to the eyes tracing the tail of the Fe-3d IB. (b) Enlarged view of Fig. 5(a), where green curves represent the Fermi-Dirac distribution curves with backgrounds. Each curve is normalized so that the value at $E_F$ of the Fermi-Dirac distribution curve is 0.5. Red and green triangles represent the points at $E_F$ of the intrinsic Fe PDOS, and Fermi-Dirac distribution curves, respectively.

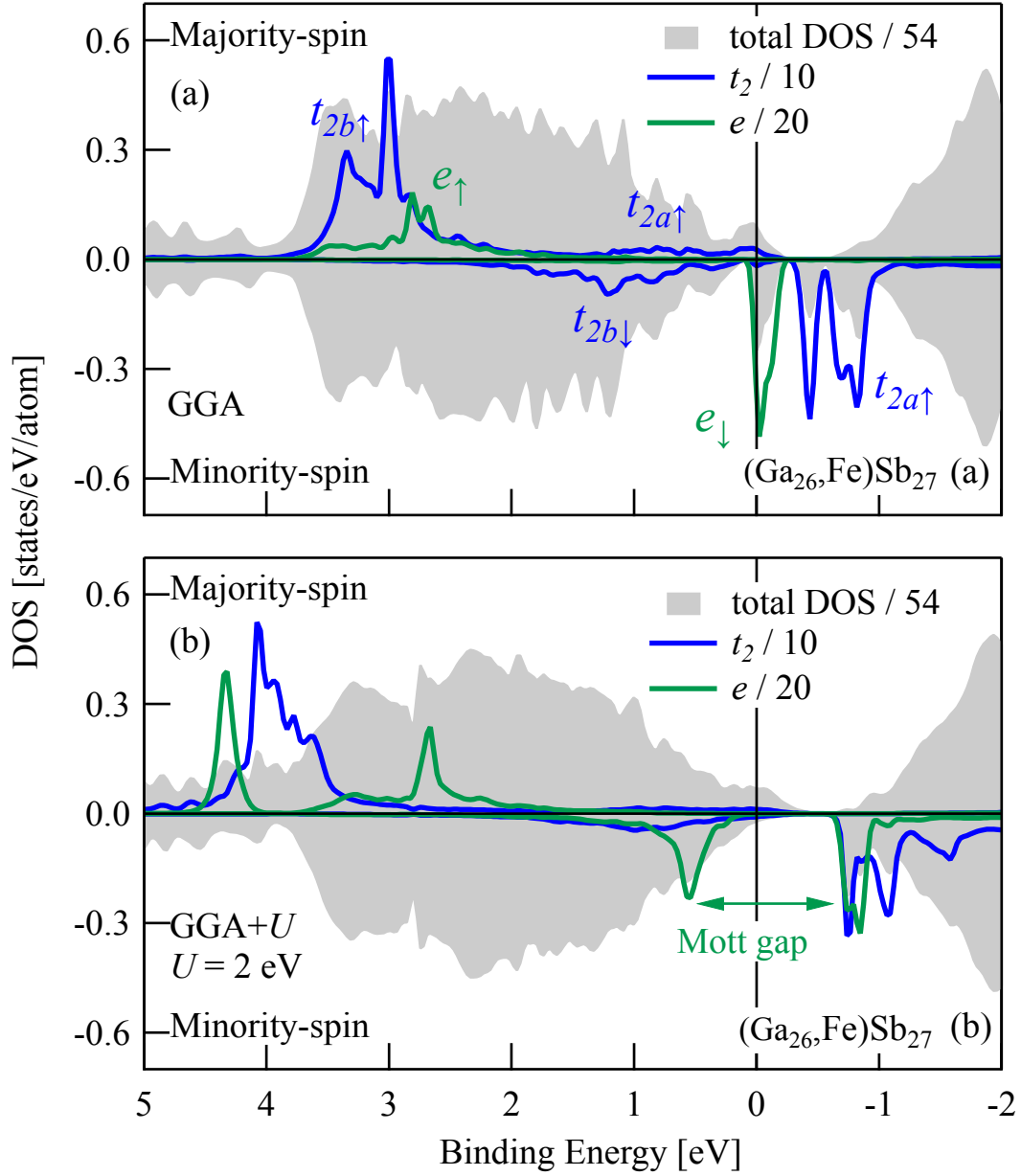

FIG. 6: Spin-resolved density of states (DOS) of a $(Ga_{26},Fe_1)Sb_{27}$ supercell calculated using GGA (a) and GGA+$U$ ($U = 2$ eV) (b) methods. The total DOS divided by the number of atoms in the supercell is plotted by gray area, the PDOS of the Fe $e$ and $t_2$ orbitals are shown by green and blue curves, respectively.

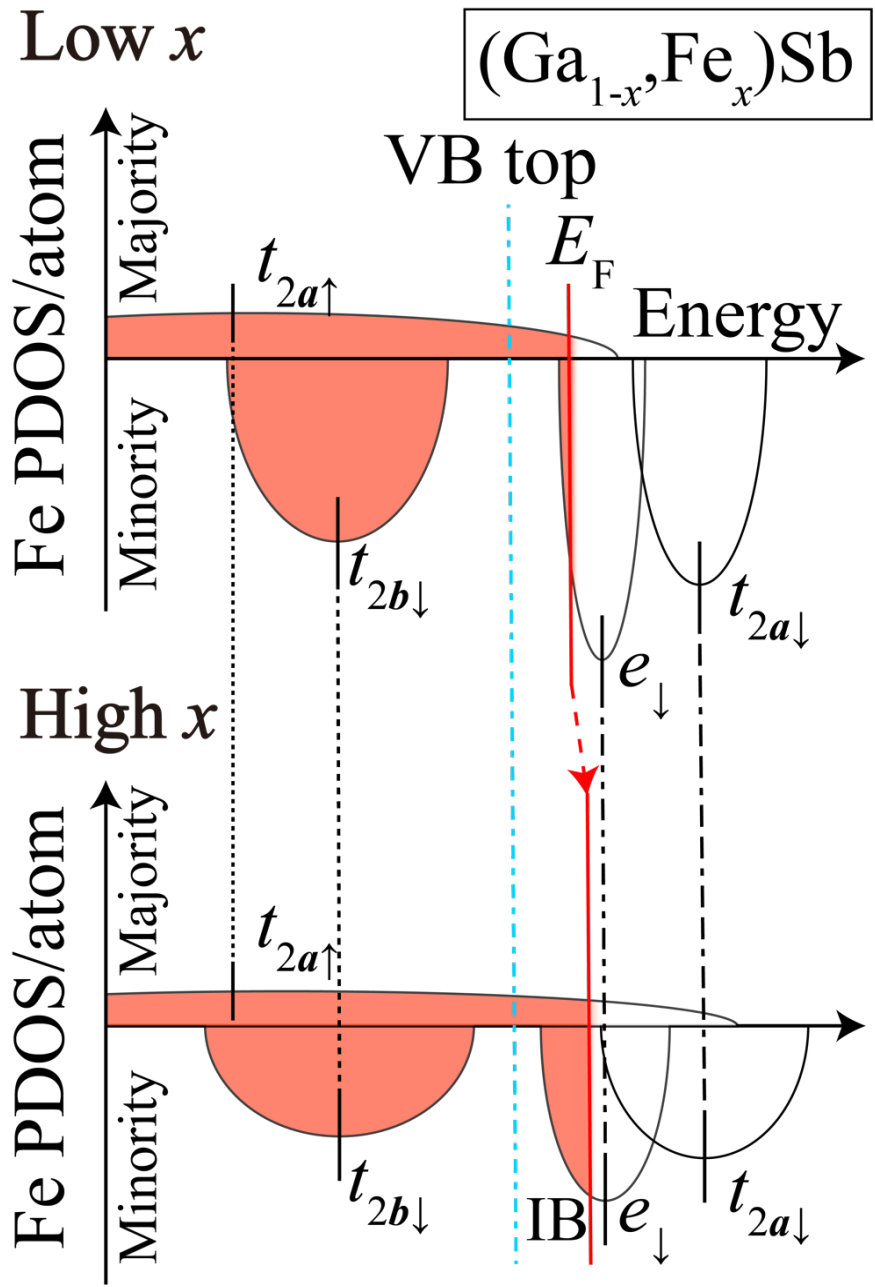

FIG. 7: Schematic energy diagram of Fe PDOS per Fe atom for the low (upper panel) and high (lower panel) Fe concentration.